\documentclass[pdflatex,sn-mathphys-num]{nature_config/sn-jnl}

\usepackage{graphicx}%
\usepackage{multirow}%
\usepackage{amsmath,amssymb,amsfonts}%
\usepackage{amsthm}%
\usepackage{mathrsfs}%
\usepackage[title]{appendix}%
\usepackage{xcolor}%
\usepackage{textcomp}%
\usepackage{manyfoot}%
\usepackage{booktabs}%
\usepackage{algorithm}%
\usepackage{algorithmicx}%
\usepackage{algpseudocode}%
\usepackage{listings}%
\usepackage{subcaption}

\theoremstyle{thmstyleone}%
\theoremstyle{thmstyletwo}%

\theoremstyle{thmstylethree}%

\begin{document}

\title[Article Title]{Global Evaluation of AI and NWP Precipitation Forecasts During Atmospheric River Events}


\author*[1,2]{\fnm{Marina} \sur{Vicens-Miquel}}\email{marinavicensmiquel@gmail.com}
\author[2,3]{\fnm{Taylor} \sur{Mandelbaum}}
\author[1,2]{\fnm{Amy} \sur{McGovern}}
\author[1,2]{\fnm{Aaron J.} \sur{Hill}}
\author[1,2]{\fnm{Daniel} \sur{Rothenberg}}

\affil*[1]{\orgdiv{School of Meteorology and School of Computer Science}, \orgname{University of Oklahoma}, \orgaddress{\city{Norman}, \postcode{73072}, \state{OK}, \country{USA}}}

\affil[2]{\orgname{NSF AI Institute for Research on Trustworthy AI in Weather, Climate, and Coastal Oceanography}, \orgaddress{\city{Norman}, \postcode{73072}, \state{OK}, \country{USA}}}

\affil[3]{\orgname{Brightband}, \orgaddress{\city{San Francisco}, \state{CA}, \country{USA}}}

\abstract{Atmospheric rivers (ARs) produce many of the world's most extreme precipitation events and hydrometeorological hazards. Although artificial intelligence weather prediction (AIWP) models have demonstrated skill comparable to or exceeding numerical weather prediction (NWP) systems for large-scale atmospheric variables, their ability to forecast AR-related precipitation remains insufficiently characterized globally. Here, we evaluate 24-hour precipitation forecasts from the Global Forecast System (GFS), Global Ensemble Forecast System (GEFS), GraphCast, and Artificial Intelligence Forecasting System (AIFS) from Day 1 through Day 10 globally and across North America, Europe, and Australia and New Zealand. Using the Extreme Weather Bench framework, forecasts are evaluated against Integrated Multi-satellitE Retrievals for GPM (IMERG) observations using measures of precipitation magnitude, spatial structure, and localization. GraphCast and AIFS exhibit greater spatial skill than GFS and GEFS, particularly for heavy precipitation and at longer lead times, and better preserve the spatial organization of AR-related precipitation through Day 10. However, this improved spatial skill does not translate into accurate precipitation magnitudes. AIWP models tend to overpredict moderate-to-heavy accumulations while underpredicting the heaviest precipitation at longer lead times, whereas NWP systems develop pronounced dry biases. These results reveal distinct strengths and limitations of AIWP for high-impact precipitation forecasting and provide a reproducible benchmark.}




\maketitle

\section{Introduction}\label{sec_introduction}

Atmospheric rivers (ARs) are long, narrow corridors of enhanced horizontal water vapor transport that play a fundamental role in the global hydrological cycle and poleward moisture transport in the midlatitudes \citep{zhu1998proposed,ralph2018defining,newell1992tropospheric}. Although they occupy only a small fraction of the Earth's atmosphere at any given time, ARs transport much of the horizontal water vapor flux outside the tropics \citep{zhu1998proposed,newell1992tropospheric}. When this moisture encounters favorable dynamical conditions, including orographic lifting, frontal systems, and extratropical cyclones, it can produce some of the most extreme rainfall and snowfall events worldwide \citep{ralph2011storms,dacre2015atmospheric,dettinger2011atmospheric}. Consequently, ARs contribute substantially to annual precipitation and water resources in regions including western North America, western Europe, Chile, and New Zealand, while also producing damaging floods, landslides, and other hydrometeorological hazards \citep{dettinger2011atmospheric,waliser2017extreme}.

Despite advances in numerical weather prediction (NWP), forecasting precipitation during AR events remains challenging \citep{martin2018evaluation,english2021evaluating}. Previous NWP studies have demonstrated meaningful predictability of AR occurrence and moisture transport from medium-range to subseasonal lead times, while also showing increasing errors in AR location and landfall with lead time and substantial regional variability in forecast skill \citep{wicketal2013arprediction,nardietal2018reforecast,deflorio2018predictionskill}. Predicting the precipitation produced by these systems presents an additional challenge. The conversion of atmospheric moisture into precipitation depends on interactions among atmospheric dynamics, terrain, cloud microphysics, and thermodynamic processes, with orographic forcing playing a particularly important role during landfalling ARs \citep{ralph2011storms,martin2018evaluation}. Errors in AR position, timing, storm evolution, or terrain representation can therefore produce substantial errors in precipitation magnitude and spatial distribution \citep{martin2018evaluation,english2021evaluating}. Many of the societal impacts associated with ARs depend not simply on whether an AR is predicted, but on where and how much precipitation it produces. Assessing precipitation directly is therefore essential for determining the practical forecast skill of weather prediction systems during AR events \citep{dettinger2011atmospheric,ralph2019scale}.

Artificial intelligence weather prediction (AIWP) has recently emerged as an alternative approach to medium-range forecasting. Models including GraphCast \citep{lam2023learning}, the Artificial Intelligence Forecasting System (AIFS) \citep{lang2024aifs}, Pangu-Weather \citep{bi2023accurate}, and FourCastNet \citep{pathak2022fourcastnet} have demonstrated skill comparable to or exceeding traditional NWP systems for many large-scale atmospheric variables \citep{bi2023accurate,lam2023learning,rasp2026weathernext}. Precipitation, however, remains particularly challenging because it emerges from nonlinear processes spanning a wide range of spatial and temporal scales \citep{sun2014use}. Recent precipitation-specific evaluations illustrate this limitation \citep{sun2014use}. GraphCast has shown favorable performance relative to conventional NWP systems across several precipitation metrics while exhibiting limitations in representing heavy precipitation \citep{radford2025comparison}, and AIFS has reduced some precipitation forecast errors relative to IFS while retaining limitations for precipitation extremes \citep{pan2025comparative}. These evaluations, however, were conducted over geographically limited domains and were not designed specifically to assess precipitation associated with ARs.

AIWP models have also been evaluated specifically for AR prediction, primarily through integrated vapor transport (IVT) and AR object characteristics. GraphCast showed lower IVT errors than the Integrated Forecasting System (IFS) over the northeastern Pacific and western North America \citep{lam2023learning}. More targeted evaluations have subsequently demonstrated that improved gridpoint skill does not necessarily translate into improved representation of AR objects. Comparisons of GraphCast and Pangu-Weather with conventional NWP systems found that AIWP models could perform better according to gridpoint metrics while exhibiting comparable or poorer object-based AR performance \citep{swensonetal2026}. Similarly, an evaluation of multiple AIWP models for ARs affecting the western United States found increasing advantages for AIWP in individual atmospheric variables at longer lead times, while IFS retained greater spatial skill for AR objects at shorter lead times \citep{davis2026AR}. Together, these findings highlight an important distinction between accurately predicting atmospheric variables and representing the location and structure of high-impact weather systems.

A corresponding question remains unresolved for the precipitation produced by ARs. Existing AIWP evaluations of ARs have largely emphasized AR detection, moisture transport, or AR object characteristics, while AIWP precipitation studies have not specifically examined a globally distributed collection of AR events. It therefore remains unclear whether improvements in large-scale atmospheric prediction translate into more accurate AR-related precipitation forecasts, particularly for heavy precipitation and at longer lead times. This study evaluates 24-hour accumulated precipitation forecasts from two operational NWP systems, the Global Forecast System (GFS) and Global Ensemble Forecast System (GEFS), and two AIWP models, GraphCast and AIFS, from Day 1 through Day 10. Using the Extreme Weather Bench (EWB) framework \citep{mcgovern2026extreme}, we identify a globally distributed collection of AR events and evaluate forecast performance globally and across three major AR-affected regions: North America, Europe, and Australia and New Zealand. The evaluation considers precipitation magnitude, spatial structure, and localization, with particular emphasis on heavy precipitation, thereby distinguishing the ability of forecasting systems to reproduce the spatial organization of AR-related precipitation from their ability to reproduce its magnitude and distribution.

Our results show that GraphCast and AIFS provide greater spatial skill than GFS and GEFS, particularly for heavy precipitation and at longer lead times, and better preserve the spatial organization of AR-related precipitation through Day 10. These improvements, however, do not translate directly into more accurate precipitation magnitudes. AIWP models tend to overpredict moderate-to-heavy precipitation while underpredicting the heaviest precipitation at longer lead times, whereas the NWP systems develop increasingly pronounced dry biases. Beyond comparing current forecasting systems, the standardized event-based evaluation implemented within EWB provides a reproducible benchmark for assessing future advances in AR precipitation prediction.

\section{Results}\label{sec_results}
Event-level clustered bootstrap analysis produced narrow 95\% confidence intervals for the aggregate verification metrics across models and forecast lead times. Because these intervals are small relative to the ranges of the metrics shown in the main figures, confidence intervals for selected lead times are presented separately in Appendix \ref{appendix_bootstrap_ci}. Details of the bootstrap procedure are provided in Section \ref{sec_methods}.

\subsection{AIWP models improve the spatial prediction of AR precipitation}
\label{sec_global_results}

\begin{figure}[t]
\begin{center}
 \noindent\includegraphics[width=\textwidth]{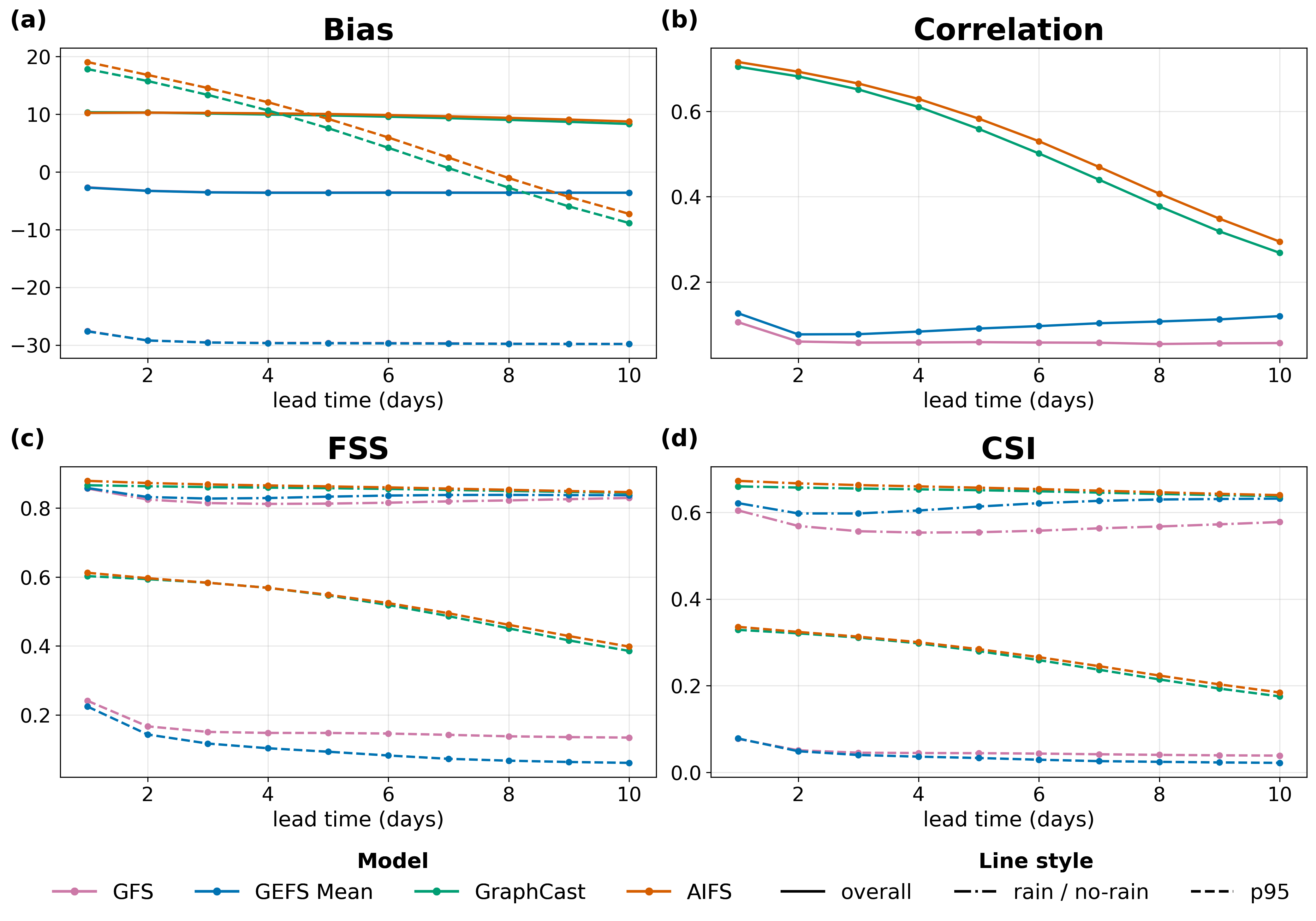}\\
\caption{Global precipitation forecast performance during atmospheric river events from Day 1 through Day 10 for GFS, GEFS Mean, GraphCast, and AIFS. Panels show (a) bias, (b) Pearson correlation, (c) Fractions Skill Score (FSS), and (d) Critical Success Index (CSI). Line styles distinguish overall, rain/no-rain, and 95th-percentile precipitation metrics.}
 \label{global_overall_perfomance}
\end{center}
\end{figure}

GraphCast and AIFS exhibit greater spatial skill than GFS and the GEFS ensemble mean across the global AR dataset, with the largest differences occurring for heavy precipitation and at longer forecast lead times (Fig.~\ref{global_overall_perfomance}). The two AIWP models perform similarly throughout the forecast range, with AIFS providing a small advantage for several metrics. Their improvement is particularly pronounced for Pearson correlation and precipitation exceeding the local 95th-percentile (p95) threshold, whereas differences among the four models are smaller for general rain/no-rain occurrence.

The model families exhibit distinct systematic errors in precipitation magnitude (Fig.~\ref{global_overall_perfomance}a). GFS and GEFS Mean show a persistent negative overall bias and substantially greater underprediction at observed p95 locations. GraphCast and AIFS instead exhibit an overall wet bias throughout the forecast range. For p95 precipitation, both AIWP models transition from substantial overprediction at short lead times toward underprediction at longer lead times, with bias approaching zero around Days 7--8 before becoming negative by Day 10. Thus, the greater spatial skill of the AIWP models is accompanied by systematic errors in precipitation magnitude that evolve with forecast lead time.

GraphCast and AIFS achieve correlations above 0.70 at Day 1, decreasing to approximately 0.27--0.29 by Day 10 (Fig.~\ref{global_overall_perfomance}b). GFS and GEFS Mean remain below approximately 0.13 throughout the forecast range. Differences are similarly pronounced for heavy precipitation. At Day 1, GraphCast and AIFS achieve p95 FSS values of approximately 0.61 compared with approximately 0.22--0.24 for GFS and GEFS Mean (Fig.~\ref{global_overall_perfomance}c). Although skill decreases with lead time, the AIWP models retain substantially higher p95 FSS and CSI through Day 10 (Fig.~\ref{global_overall_perfomance}c,d). In contrast, differences among models are considerably smaller for rain/no-rain occurrence. The AIWP advantage is therefore strongest for the spatial prediction of heavy AR-related precipitation rather than for distinguishing broadly between raining and non-raining areas.

\begin{figure}[t]
    \centering
    \includegraphics[width=\textwidth]{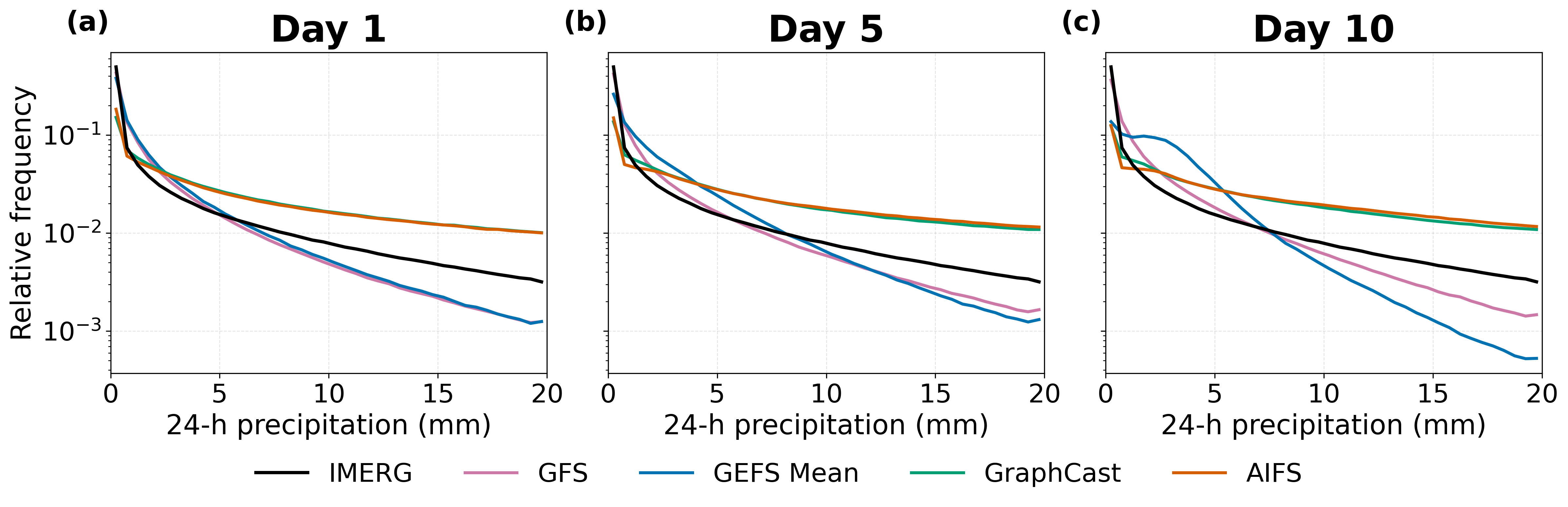}
    \caption{Global distributions of 24-hour precipitation during atmospheric river events at (a) Day 1, (b) Day 5, and (c) Day 10 for IMERG, GFS, GEFS Mean, GraphCast, and AIFS. Curves show the relative frequency of precipitation amounts from 0 to 20~mm and are averaged across all available 24-hour AR evaluation periods with each period weighted equally. The IMERG curve represents the mean of the observational distributions aligned to the native grids of the four forecast systems. Note the logarithmic scale on the vertical axis.}
    \label{global_pdf}
\end{figure}

The distributions of 24-hour precipitation accumulations reveal a different aspect of forecast performance (Fig.~\ref{global_pdf}). At Day 1, GFS and GEFS Mean closely follow Integrated Multi-satellitE Retrievals for Global Precipitation Measurement (IMERG) across much of the evaluated distribution, whereas GraphCast and AIFS produce moderate-to-heavy precipitation more frequently than observed. With increasing lead time, GFS and GEFS Mean increasingly underrepresent higher precipitation amounts, while the AIWP models retain distributions shifted toward more frequent moderate-to-heavy precipitation. By Day 10, the separation between the two model families is particularly pronounced (Fig.~\ref{global_pdf}c). These results demonstrate that greater spatial skill does not necessarily imply a more accurate precipitation amount distribution. GraphCast and AIFS better reproduce where heavy AR-related precipitation occurs while retaining systematic errors in its frequency and magnitude.

\subsection{Forecast differences persist across regions and seasons}
\label{sec_regional_results}

\begin{figure}[t]
    \centering
    \includegraphics[width=\textwidth]{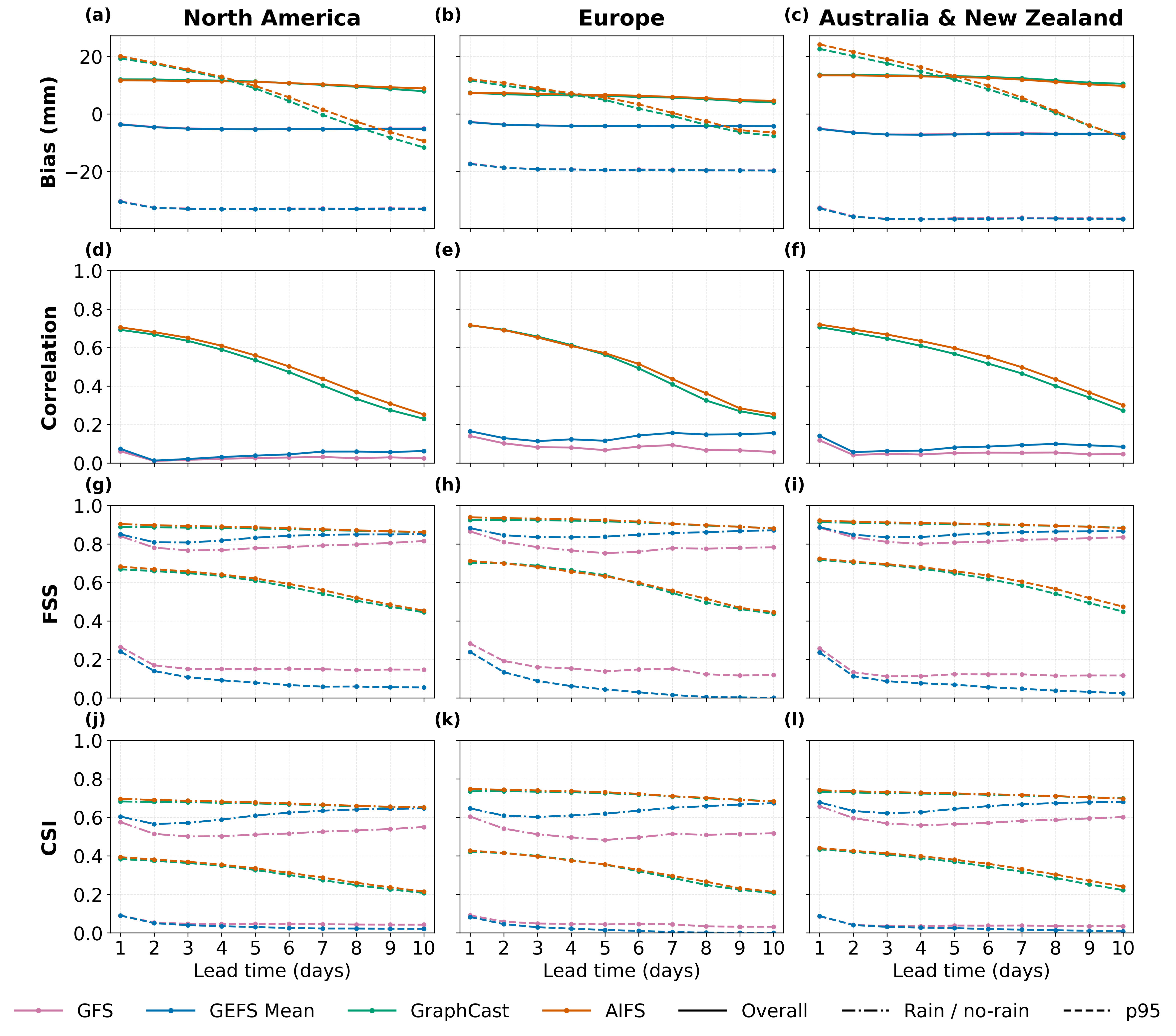}
    \caption{Regional precipitation forecast performance during atmospheric river events from Day 1 through Day 10 for GFS, GEFS Mean, GraphCast, and AIFS across North America (a, d, g, j), Europe (b, e, h, k), and Australia and New Zealand (c, f, i, l). Rows show bias (a–c), Pearson correlation (d–f), Fractions Skill Score (FSS; g–i), and Critical Success Index (CSI; j–l). Bias is shown for the overall precipitation field and the 95th-percentile (p95) threshold, while FSS and CSI are evaluated for rain/no-rain occurrence and p95 precipitation. Metrics are averaged across all 24-hour AR evaluation periods within each region, with each evaluation period weighted equally.}
    \label{regional_overall_eval}
\end{figure}

The global differences between AIWP and NWP systems persist across North America, Europe, and Australia and New Zealand (Fig.~\ref{regional_overall_eval}). GraphCast and AIFS consistently exhibit higher spatial correlation than GFS and GEFS Mean across all three regions, although skill decreases with forecast lead time (Fig.~\ref{regional_overall_eval}d--f). The separation is again strongest for heavy precipitation, with the AIWP models maintaining substantially higher p95 FSS and CSI throughout the 10-day forecast period (Fig.~\ref{regional_overall_eval}g--l). Differences among models are smaller for rain/no-rain occurrence, consistent with the global evaluation.

Systematic precipitation biases are also consistent across regions (Fig.~\ref{regional_overall_eval}a--c). GraphCast and AIFS generally exhibit positive overall biases, whereas GFS and GEFS Mean predominantly underpredict precipitation. For p95 precipitation, the AIWP models transition from overprediction at shorter lead times toward underprediction at longer lead times, although the timing and magnitude of this transition vary among regions. GFS and GEFS Mean exhibit persistent negative p95 biases. The consistency of these characteristics indicates that the principal differences between AIWP and NWP systems are not driven by a single geographical region.

\begin{figure}[!htbp]
    \centering
    \includegraphics[width=\textwidth]{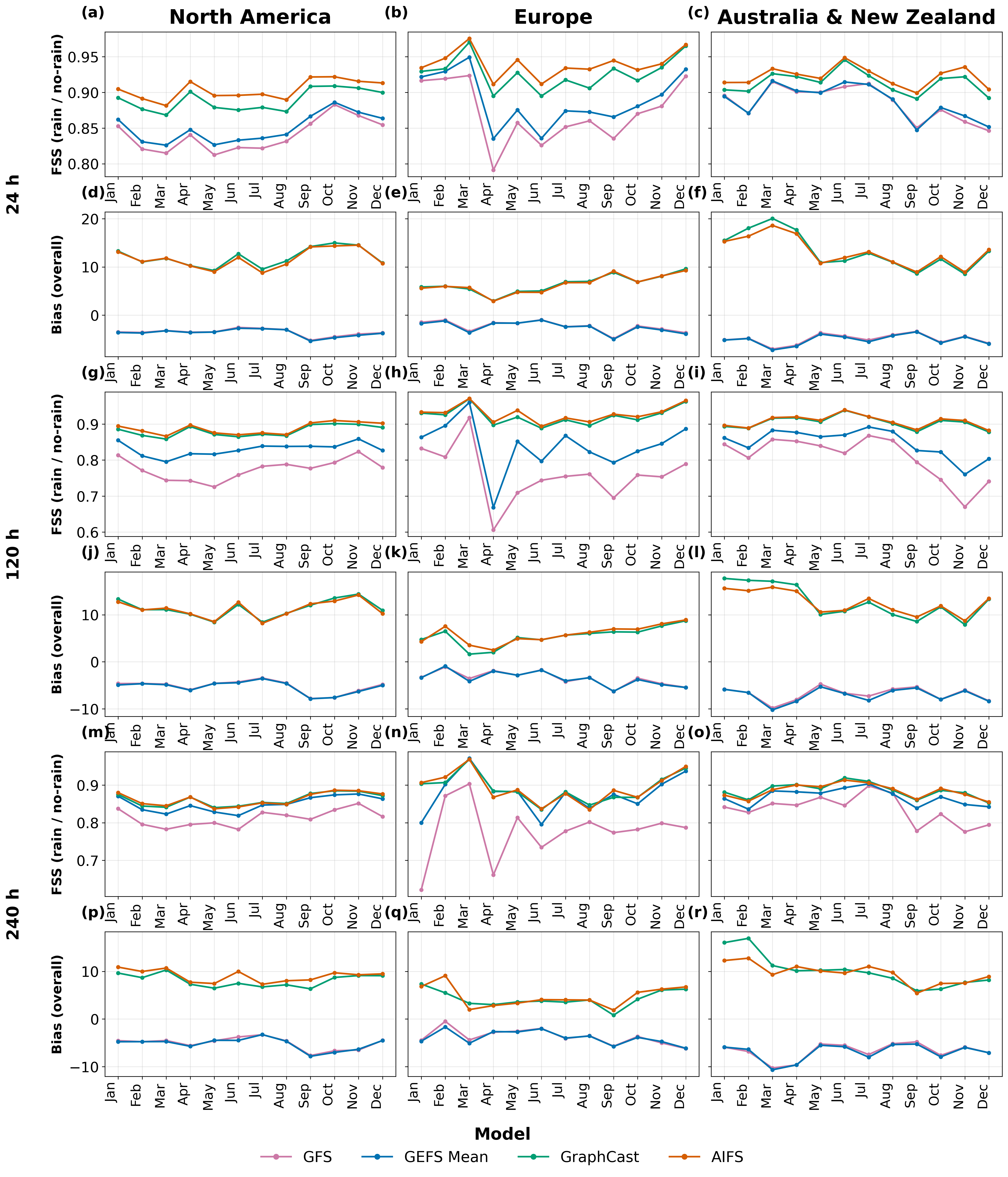}
    \caption{Seasonal variability in regional atmospheric river precipitation forecast performance for GFS, GEFS Mean, GraphCast, and AIFS. Monthly mean rain/no-rain Fractions Skill Score (FSS) and overall precipitation bias are shown for Day 1 (24 h; a–f), Day 5 (120 h; g–l), and Day 10 (240 h; m–r) forecasts across North America (a, d, g, j, m, p), Europe (b, e, h, k, n, q), and Australia and New Zealand (c, f, i, l, o, r). Within each lead time, the upper row shows FSS and the lower row shows bias. Metrics are averaged across all AR evaluation periods occurring within each calendar month.}
    \label{regional_seasonality}
\end{figure}

Seasonal variability modifies the magnitude, but not the overall pattern, of these differences (Fig.~\ref{regional_seasonality}). North America exhibits comparatively modest seasonal variations in FSS, particularly for GraphCast and AIFS, whereas Europe shows stronger month-to-month variability, especially for GFS and GEFS Mean. Australia and New Zealand also exhibit pronounced seasonal variations with a different seasonal cycle from the Northern Hemisphere regions. Bias varies similarly with region and season, with the largest positive AIWP biases generally occurring in Australia and New Zealand and smaller biases in Europe. Despite this variability, GraphCast and AIFS generally maintain higher rain/no-rain FSS than the NWP systems across seasons and lead times, indicating that their relative spatial advantage is robust across diverse AR environments.

\subsection{AIWP models preserve large-scale precipitation structure}
\label{sec_case_studies}

\begin{figure}[t]
    \centering
    \includegraphics[width=\textwidth]{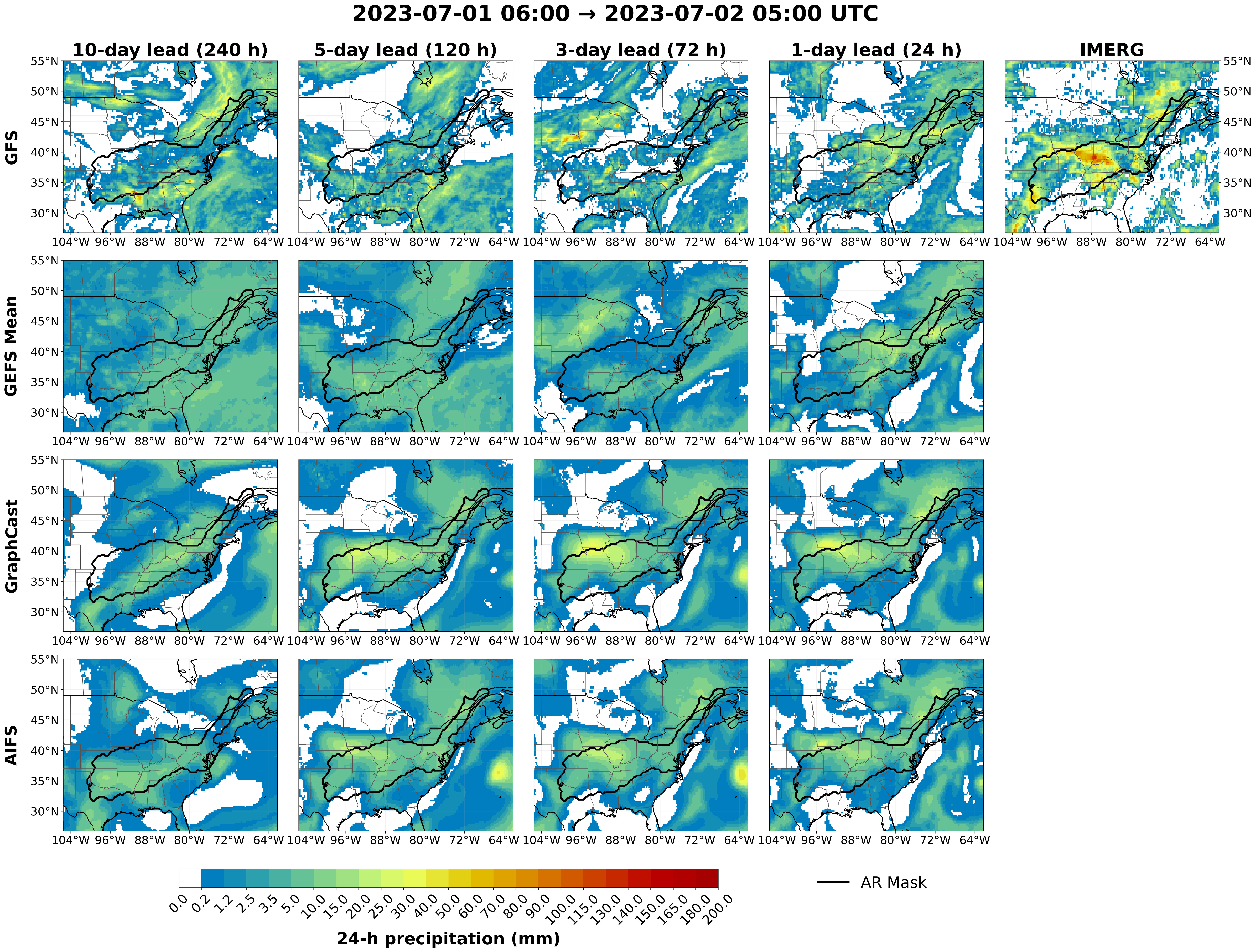}
    \caption{Representative North American atmospheric river event valid from 1 July 2023 at 06:00 UTC to 2 July 2023 at 05:00 UTC. IMERG 24-hour accumulated precipitation is compared with GFS, GEFS Mean, GraphCast, and AIFS forecasts at 10-, 5-, 3-, and 1-day lead times. The black contour indicates the AR mask used to define the event.}
    \label{north_america_ar_event}
\end{figure}

\begin{figure}[t]
    \centering
    \includegraphics[width=\textwidth]{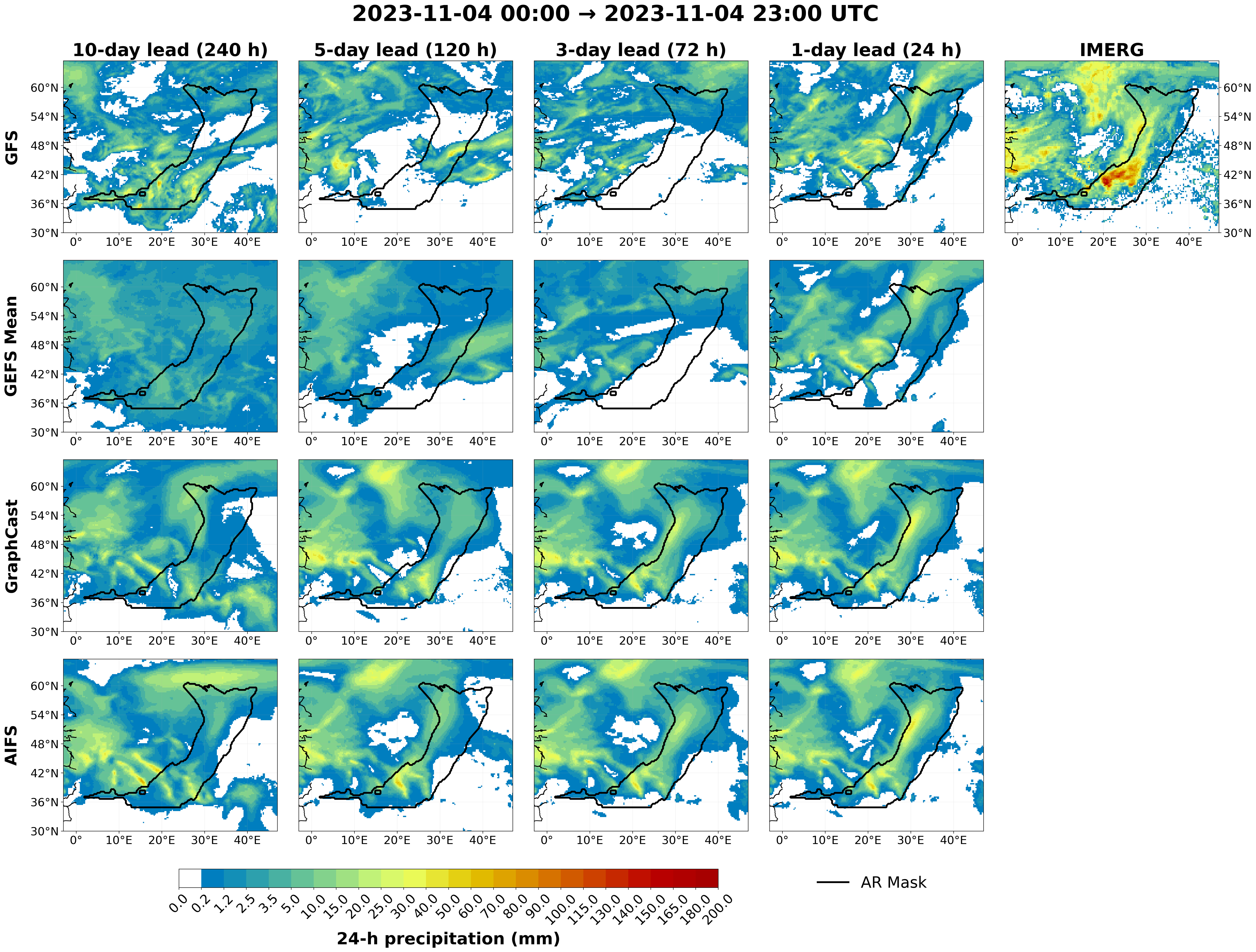}
    \caption{As in Fig.~\ref{north_america_ar_event}, but for a representative European atmospheric river event valid from 4 November 2023 at 00:00 UTC to 23:00 UTC.}
    \label{europe_ar_event}
\end{figure}

\begin{figure}[t]
    \centering
    \includegraphics[width=\textwidth]{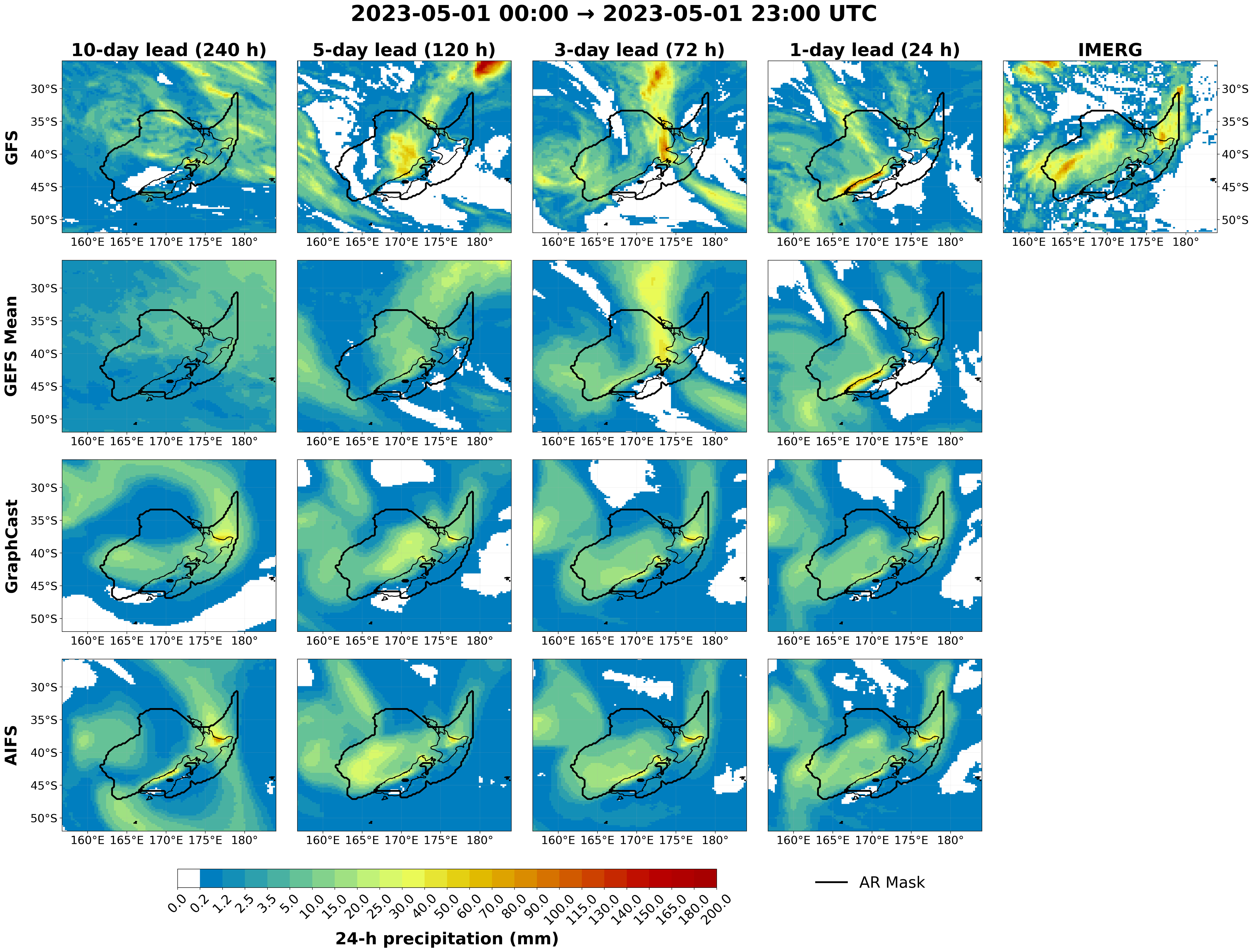}
    \caption{As in Fig.~\ref{north_america_ar_event}, but for a representative Australia and New Zealand atmospheric river event valid from 1 May 2023 at 00:00 UTC to 23:00 UTC.}
    \label{aus_nz_ar_event}
\end{figure}

Representative AR events from North America, Europe, and Australia and New Zealand illustrate the spatial characteristics underlying the aggregate verification results (Figs.~\ref{north_america_ar_event}--\ref{aus_nz_ar_event}). Across the three events, GraphCast and AIFS generally produce smoother, more spatially coherent precipitation fields that preserve the broad organization of AR-related precipitation across multiple lead times. GFS and GEFS Mean retain greater fine-scale variability but exhibit larger changes in precipitation placement and structure, particularly at longer lead times. None of the forecasting systems consistently reproduces the localized structure and magnitude of the observed precipitation maxima.

The North American event illustrates these differences particularly clearly (Fig.~\ref{north_america_ar_event}). IMERG shows an elongated precipitation structure extending from the central toward the northeastern United States, with the largest accumulations concentrated in the southwestern portion of the AR domain. At shorter lead times, all four models reproduce its broad orientation, whereas differences become more pronounced with increasing lead time. GraphCast and AIFS retain a coherent large-scale precipitation feature, while the NWP forecasts exhibit greater displacement and structural differences from the observations.

Similar characteristics occur in the European and Australia and New Zealand events (Figs.~\ref{europe_ar_event},~\ref{aus_nz_ar_event}). The AIWP models broadly reproduce the large-scale organization of AR-related precipitation but generate smoother fields and do not consistently capture localized observed maxima. The NWP systems retain greater small-scale structure but show larger displacement from observed precipitation in these examples, particularly at longer lead times. Such displacement also affects the interpretation of percentile-conditioned bias because p95 bias is evaluated at locations where observed precipitation exceeds the local threshold, displacement of forecast precipitation away from an observed maximum can produce negative bias even when comparable precipitation occurs nearby.

These examples therefore reinforce the distinction between spatial skill and precipitation magnitude identified in the aggregate analysis. AIWP models can better preserve the large-scale location and organization of AR-related precipitation while simultaneously overrepresenting moderate-to-heavy accumulations in the overall precipitation distribution and failing to reproduce individual observed maxima. Improved spatial prediction does not necessarily imply improved representation of precipitation magnitude or localized extremes.

\section{Discussion}\label{sec_discussion}

This global evaluation reveals a consistent distinction between the spatial and magnitude characteristics of AR precipitation forecasts produced by AIWP and operational NWP systems. GraphCast and AIFS more accurately reproduce the spatial organization of AR-related precipitation than GFS and GEFS Mean, with the largest improvements occurring for heavy precipitation and at longer forecast lead times. This advantage persists across geographically and climatologically distinct AR environments and is evident in correlation, FSS, and CSI. However, greater spatial skill does not translate directly into more accurate precipitation magnitudes. The AIWP models systematically overrepresent moderate-to-heavy precipitation while progressively underpredicting the heaviest observed accumulations at longer lead times, whereas GFS and GEFS Mean exhibit pronounced dry biases. Evaluating forecast quality using spatial metrics alone would therefore provide an incomplete characterization of AIWP performance for high-impact precipitation.

The strong spatial performance of AIWP models is consistent with their demonstrated ability to predict large-scale atmospheric circulation and moisture transport \citep{lam2023learning,bi2023accurate}. Previous AR-specific evaluations have nevertheless shown that improvements in gridpoint atmospheric variables do not necessarily translate into improved representation of AR objects \citep{swensonetal2026,davis2026AR}. Our results extend this distinction to the precipitation produced by ARs. GraphCast and AIFS preserve the broad organization and location of AR-related precipitation more effectively than the NWP systems evaluated here, including at longer lead times, yet retain systematic errors in the precipitation amount distribution and localized extremes. Skill in predicting the atmospheric environment of a high-impact event and skill in predicting its hydrological expression should therefore be considered related but distinct forecasting capabilities.

The precipitation-magnitude errors identified here are also consistent with broader evaluations showing persistent challenges in AIWP precipitation forecasts \citep{radford2025comparison,pan2025comparative}. Precipitation depends on processes operating across scales that are not fully resolved by global forecasting systems, including cloud microphysics, convection, orographic interactions, and local thermodynamic variability \citep{sun2014use}. The smoother precipitation structures evident in GraphCast and AIFS may contribute to their greater large-scale spatial correspondence while limiting their representation of localized maxima. Conversely, the greater fine-scale variability of the NWP forecasts does not necessarily correspond to greater spatial accuracy, particularly when precipitation features are displaced from their observed locations. These results emphasize the importance of evaluating both spatial organization and precipitation magnitude when assessing emerging forecasting systems.

The consistency of the AIWP spatial advantage across North America, Europe, and Australia and New Zealand indicates that the global results are not attributable to a single AR regime. Nevertheless, regional and seasonal variations in bias and skill demonstrate that forecast performance remains sensitive to environmental and geographical context. Orographic forcing, storm-track characteristics, moisture availability, and seasonal circulation differences can alter the mechanisms controlling AR precipitation and may contribute to this variability. Further work examining performance across specific AR magnitudes and meteorological regimes could help determine where AIWP improvements are most pronounced and where systematic limitations remain.

Several considerations affect the interpretation of these results. IMERG provides the globally consistent observational reference required for this analysis, but satellite precipitation estimates remain uncertain, particularly for heavy precipitation and complex terrain \citep{derin2022evaluation}. AR identification is similarly dependent on the detection methodology; different algorithms can produce differences in AR frequency, duration, spatial extent, and associated precipitation \citep{shields2018atmospheric,rutz2019atmospheric,o2022increases}. Here, the same EWB AR definition and observational reference are applied to every forecasting system, providing a controlled basis for relative model comparison, although sensitivity to alternative observational products and AR definitions remains an important direction for future evaluation.

Persistent ARs can also contribute multiple overlapping 24-hour evaluation periods. We retain these windows to evaluate forecast performance throughout long-duration events rather than selecting a single period from each AR. Their dependence is explicitly accounted for in the uncertainty analysis through event-level clustered bootstrap resampling, in which all windows associated with a continuous AR are resampled together. The resulting confidence intervals remain narrow relative to the differences among forecasting systems, indicating that the principal aggregate results are robust to sampling variability among AR events.

Together, these findings demonstrate that recent AIWP systems provide meaningful improvements in the spatial prediction of AR-related precipitation but have not resolved the challenge of accurately representing precipitation magnitude and localized extremes. This distinction is particularly important for high-impact forecasting, where correctly locating a precipitation system and accurately predicting how much precipitation it produces have different implications for flood forecasting, water-resource management, and hazard preparedness. The event-based evaluation developed here provides a reproducible framework for tracking both aspects of forecast performance as AI and physics-based weather prediction systems continue to evolve.

\section{Methods}\label{sec_methods}
The evaluation framework consists of identifying and tracking AR events, constructing 24-hour precipitation evaluation windows, and assessing forecast performance across multiple forecast lead times. The same framework is applied to all forecasting systems to enable consistent comparisons globally and across the three regional domains.

\subsection{Datasets and forecast systems}
AR events were identified using atmospheric fields from the fifth-generation European Centre for Medium-Range Weather Forecasts (ECMWF) atmospheric reanalysis (ERA5) \citep{hersbach2020era5}, and IMERG observations \citep{huffman2020integrated} were used as the precipitation reference. ERA5 provides hourly global atmospheric fields at 0.25$^{\circ}$ resolution, including the specific humidity and zonal and meridional wind fields required for AR identification. IMERG provides satellite-based precipitation estimates at 0.10$^{\circ}$ spatial resolution.

Precipitation forecasts were obtained from four global forecasting systems. The GFS \citep{NOAA_GFS} and GEFS ensemble mean \citep{zhou2022development} represent operational NWP systems, while GraphCast \citep{lam2023learning} and the deterministic AIFS \citep{lang2024aifs} represent AIWP models. GFS and GEFS forecasts were obtained through the NOAA Open Data Dissemination program via the Registry of Open Data on Amazon Web Services \citep{NOAA_GEFS_AWS,NOAA_GFS_AWS}. GraphCast and AIFS forecasts were obtained from the same forecast archive used by \citet{mcgovern2026extreme} and were initialized from Integrated Forecasting System high-resolution (IFS HRES). All four forecast datasets provide precipitation fields at 0.25$^{\circ}$ spatial and 6-h temporal resolution for the analysis conducted here.

To ensure direct comparison among forecasting systems, all datasets were restricted to the common period from April 2021 through December 2024. Twenty-four-hour precipitation accumulations were constructed by aggregating the hourly IMERG observations and the 6-hour precipitation accumulations from each forecasting system. Forecasts were evaluated from Day 1 through Day 10.

IMERG was spatially coarsened and aligned separately to each forecast grid while preserving the native forecast fields. Aggregation factors were determined from the ratio between the forecast and IMERG grid spacings, and the maximum IMERG precipitation value within each model-scale spatial block was retained. Maximum-based aggregation was used to preserve localized heavy-precipitation features that could otherwise be smoothed by spatial averaging \citep{schwartz2017comparison}. The resulting observational field was aligned to the corresponding forecast grid using nearest-neighbor reindexing. The forecast fields themselves were therefore not spatially interpolated, and all verification metrics were calculated on their native grids.

\subsection{Atmospheric river identification and event tracking}
\label{sec:ar_identification}

ARs were identified globally using the detection methodology implemented in EWB \citep{EWB_GitHub,mcgovern2026extreme}. The EWB algorithm was applied to hourly ERA5 specific humidity and zonal and meridional wind fields on a 0.25$^{\circ}$ grid. Integrated vapor transport (IVT) was calculated from vertically integrated zonal and meridional moisture fluxes and processed using a Gaussian-smoothed Laplacian ($\sigma=3$). Using the default EWB criteria, candidate ARs were defined by an IVT threshold of 400~kg~m$^{-1}$~s$^{-1}$ and an absolute Laplacian threshold of 2.5~kg~m$^{-2}$~s$^{-2}$. The Laplacian mask was dilated by eight grid cells and intersected with the IVT threshold mask, after which connected features smaller than 500 grid cells were removed. Features with a mean latitude within 15$^{\circ}$ of the equator were excluded. These criteria follow the EWB implementation, which builds on established AR identification approaches \citep{newell1992tropospheric,ullrich2021tempestextremes}.

\begin{figure}[!htbp]
\centering

\begin{subfigure}{\textwidth}
    \centering
    \includegraphics[width=\textwidth]{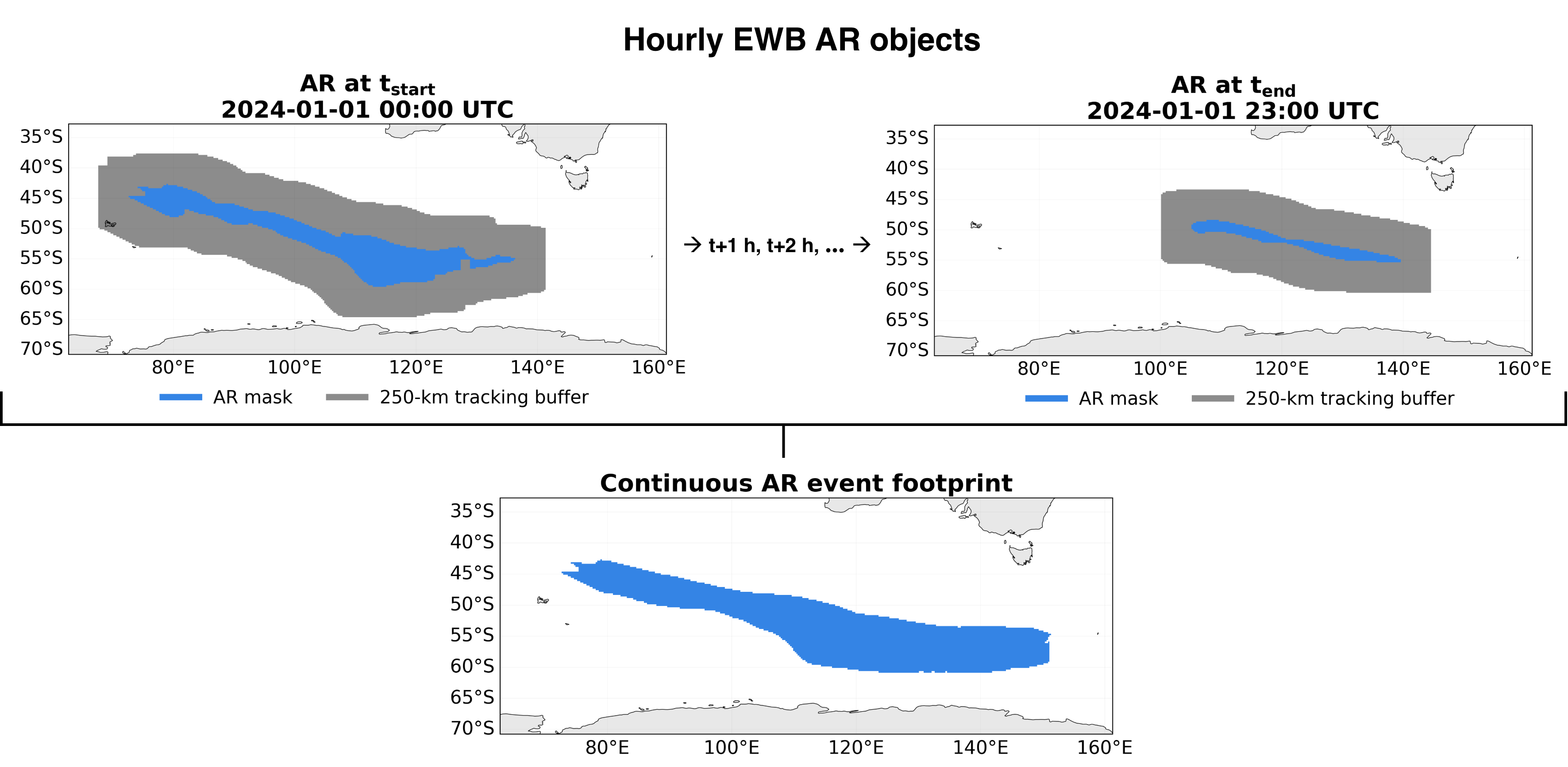}
    \caption{Hourly AR detections and track construction}
    \label{fig:ar_identification_a}
\end{subfigure}

\vspace{0.3cm}

\begin{subfigure}{\textwidth}
    \centering
    \includegraphics[width=\textwidth]{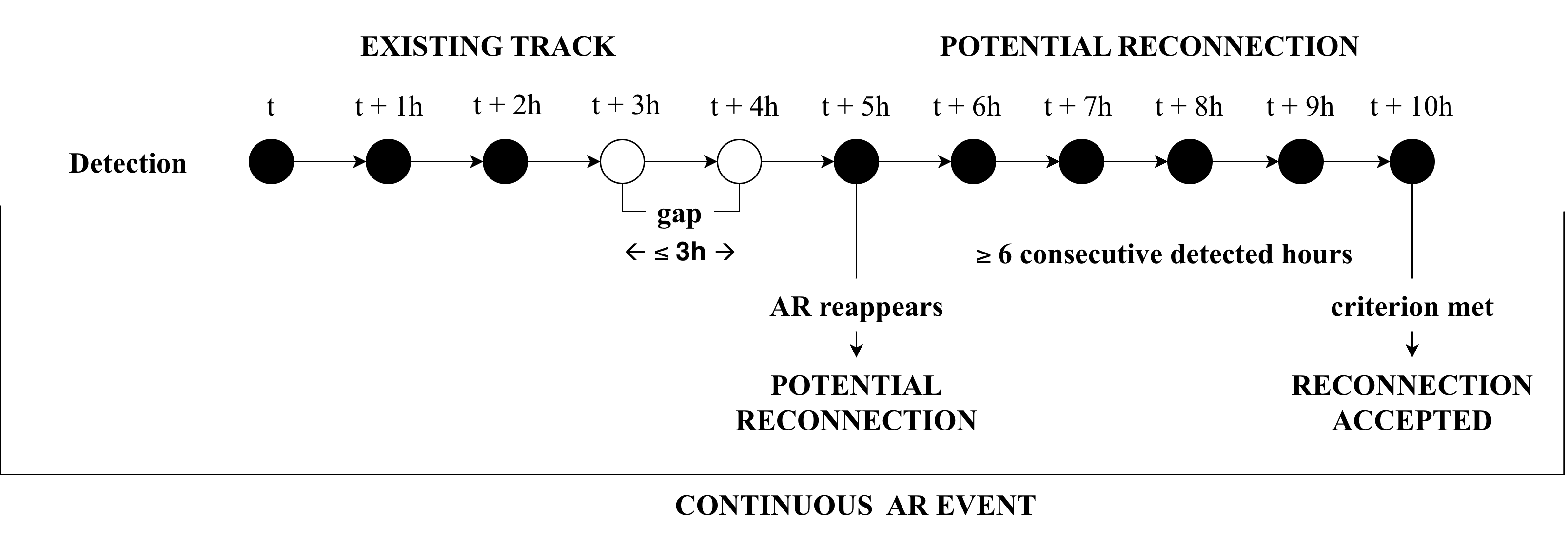}
    \caption{Temporal tracking and gap handling}
    \label{fig:ar_identification_b}
\end{subfigure}

\caption{
Schematic of the AR event construction procedure.
\textbf{a}, Hourly AR objects identified by EWB are linked through time to construct continuous AR events. Each AR mask (blue) is expanded by a 250-km tracking buffer (gray), and objects at the subsequent timestep are linked when they overlap the buffered mask. The union of the linked hourly detections defines the continuous AR event footprint. The tracking buffer is distinct from the buffer used for precipitation evaluation.
\textbf{b}, Temporal gap-handling procedure. Tracks may remain open for up to 3 consecutive hours without an AR detection. A spatially matched AR reappearing within this period is reconnected to the original track only when the new segment persists for at least 6 consecutive detected hours.
}
\label{fig:ar_identification_framework}
\end{figure}

Because EWB identifies AR objects independently at each hourly timestep, the hourly detections were linked to construct continuous AR events using the tracking procedure illustrated in Fig.~\ref{fig:ar_identification_framework}. For each existing track, the most recently matched AR mask was expanded by a 250-km buffer and compared with objects detected at the subsequent timestep (Fig.~\ref{fig:ar_identification_framework}a). Objects overlapping the buffered mask by at least one grid cell were considered candidate continuations. When multiple associations were possible, track--object pairs were assigned sequentially from largest to smallest spatial overlap, with each track and object assigned at most once per timestep. To prevent brief interruptions in AR detection from artificially splitting individual events, tracks were allowed to remain open for up to three consecutive hours without a detection (Fig.~\ref{fig:ar_identification_framework}b). During such gaps, subsequent detections were compared with the last successfully matched AR mask. A spatially matched detection reappearing within this tolerance was reconnected to the original event only when the new segment persisted for at least six consecutive detected hours. Shorter reappearances were treated as separate event candidates.

Only hours containing a detected AR contributed to event duration, and events were retained when they contained at least 24 detected AR hours. The permitted detection gaps were used only to preserve the identity of a continuous AR event when the detection algorithm briefly failed to identify the feature. No AR mask was available during these gaps, and missing masks were not interpolated or otherwise approximated. Because the hourly AR masks are subsequently combined to define the spatial domain for precipitation verification, 24-h evaluation windows were required to contain an AR detection at every hourly timestep and therefore could not span a detection gap. Event start and end times corresponded to the first and last detected AR hours. All qualifying events were retained for the global evaluation irrespective of landfall, and the first timestep at which the AR mask intersected land was recorded for the regional analyses.

\subsection{Precipitation evaluation framework}
\label{sec:ar_evaluation_windows}

\begin{figure}[t]
\centering

\begin{subfigure}{\textwidth}
    \centering
    \includegraphics[width=0.55\textwidth]{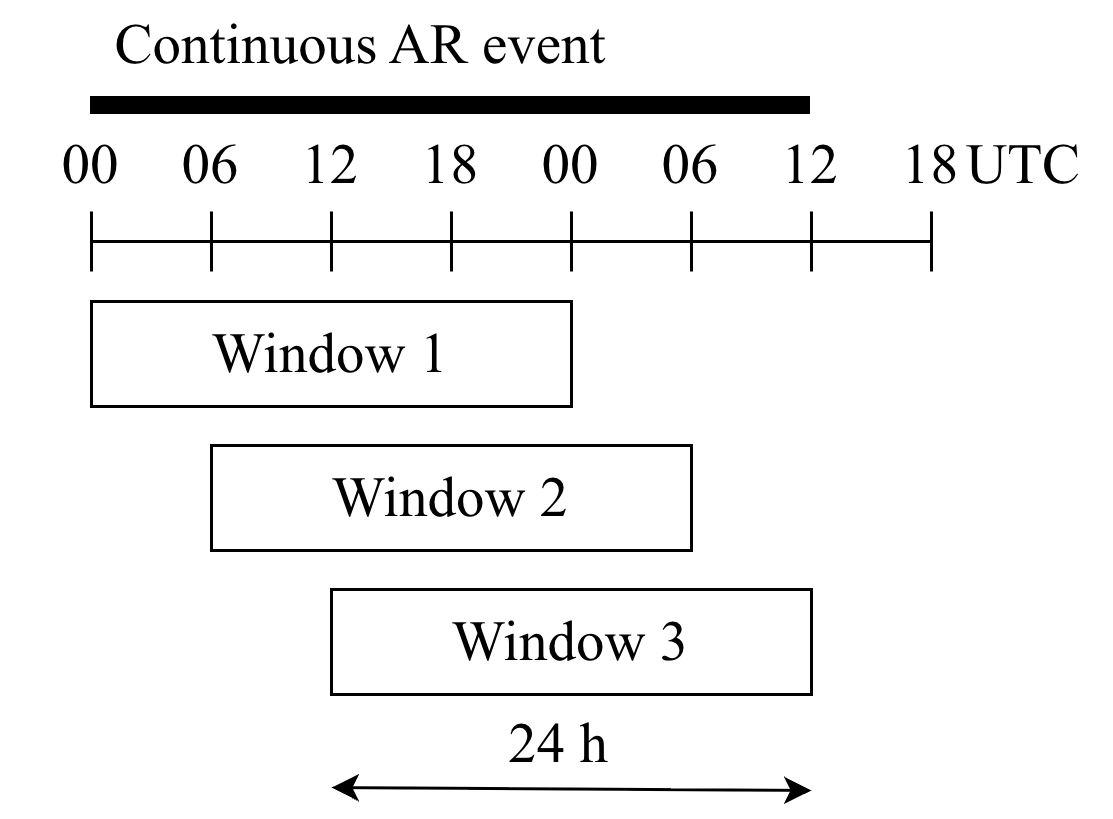}
    \caption{Temporal construction of evaluation windows}
    \label{fig:ar_windows_a}
\end{subfigure}

\vspace{0.3cm}

\begin{subfigure}{\textwidth}
    \centering
    \includegraphics[width=\textwidth]{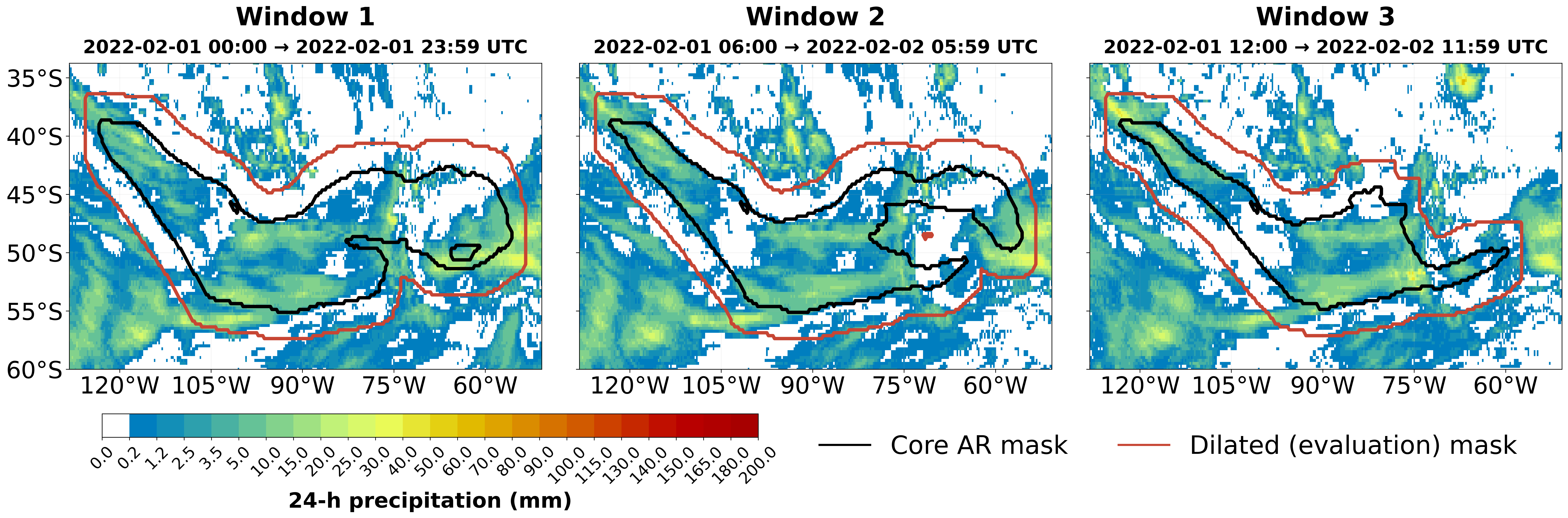}
    \caption{Spatial construction of evaluation domains}
    \label{fig:ar_windows_b}
\end{subfigure}

\caption{
Schematic of the AR precipitation evaluation framework.
\textbf{a}, Continuous AR events are divided into 24-h evaluation windows beginning at 00, 06, 12, and 18 UTC. For ARs persisting longer than 24 h, windows are generated at 6-h intervals, resulting in an 18-h overlap. Only windows with AR detections during all 24 consecutive hours are retained.
\textbf{b}, Spatial evaluation domains for three consecutive 24-h windows. Hourly AR masks are combined to define the AR footprint for each window (black), and a 250-km buffer is applied to define the precipitation evaluation domain (red).
}
\label{fig:ar_evaluation_framework}

\end{figure}

Continuous AR events were divided into 24-h windows for precipitation verification (Fig.~\ref{fig:ar_evaluation_framework}a). Window start times were restricted to 00, 06, 12, and 18 UTC to align with forecast initialization cycles. Although brief detection gaps were allowed to preserve event continuity during tracking, evaluation windows required an AR detection during all 24 consecutive hours because gaps lack the AR masks needed to define the precipitation evaluation domain.

For ARs persisting longer than 24 h, eligible windows were generated at 6-h intervals, resulting in an 18-h overlap between consecutive windows. This approach samples precipitation forecast performance throughout persistent AR events while maintaining alignment with the forecast initialization cycles. Multiple windows originating from the same continuous AR event were therefore not treated as independent events in the statistical uncertainty analysis. Once the 24-h evaluation windows were defined, the 24 hourly AR masks within each window were combined to define its spatial AR footprint (Fig.~\ref{fig:ar_evaluation_framework}b).

Different event-selection criteria were applied to the global and regional analyses. The global analysis included all qualifying AR events irrespective of landfall. Regional analyses were conducted over North America (15--85$^{\circ}$N, 170$^{\circ}$W--50$^{\circ}$W), Europe (35--78$^{\circ}$N, 12$^{\circ}$W--69$^{\circ}$E), and Australia and New Zealand (48$^{\circ}$S--10$^{\circ}$S, 111--179$^{\circ}$E), and included only events whose first landfall occurred within the corresponding domain. Evaluation windows began at or after landfall and required the AR to remain present within the region throughout the 24-h period, ensuring that an AR mask was available for the full evaluation window. The resulting numbers of unique AR events and 24-h evaluation windows for each domain are summarized in Table~\ref{tab:ar_event_counts}.

\begin{table}[ht]
\centering
\caption{Number of unique atmospheric river (AR) events and 24-hour precipitation evaluation windows included in the global and regional analyses.}
\label{tab:ar_event_counts}

\begin{tabular}{ccc}
\hline
\textbf{Evaluation Domain} & \textbf{Unique AR Events} & \textbf{24-h Evaluation Windows} \\
\hline
Global                    & 3,564 & 46,140  \\
North America             & 566   & 5,661   \\
Europe                    & 175   & 842     \\
Australia and New Zealand & 329   & 2,652   \\
\hline
\end{tabular}

\end{table}

\subsection{Verification metrics}
\label{sec:verification_metrics}

Precipitation forecast performance was comprehensively evaluated in terms of magnitude, spatial agreement, categorical occurrence, neighborhood spatial skill, and precipitation amount distribution. Metrics included mean bias, Pearson correlation, Critical Success Index (CSI), Fractions Skill Score (FSS), and a histogram-based probability density function (PDF) distance.

Threshold-dependent metrics were evaluated using spatially varying thresholds corresponding to the 75th, 90th, 95th, and 99th percentiles (p75, p90, p95, and p99) of the IMERG 24-h precipitation climatology. Percentiles were calculated independently at each grid point and aligned to the native grid of each forecast system using the observational alignment procedure described above.

Bias was used to quantify systematic errors in precipitation magnitude and was calculated as

\begin{equation}
\mathrm{Bias}
=
\frac{1}{N}
\sum_{i=1}^{N}
\left(F_i-O_i\right),
\end{equation}

where $F_i$ and $O_i$ denote forecast and observed 24-h precipitation accumulations, respectively, and $N$ is the number of valid grid points. Positive values indicate overprediction and negative values indicate underprediction. 

Spatial agreement between forecast and observed precipitation was quantified using the Pearson correlation coefficient,

\begin{equation}
r =
\frac{
\sum_{i=1}^{N}
\left(F_i-\overline{F}\right)
\left(O_i-\overline{O}\right)
}{
\sqrt{
\sum_{i=1}^{N}
\left(F_i-\overline{F}\right)^2
}
\sqrt{
\sum_{i=1}^{N}
\left(O_i-\overline{O}\right)^2
}
},
\end{equation}

where $\overline{F}$ and $\overline{O}$ denote the spatial means of the forecast and observed precipitation fields, respectively.

Threshold exceedance was evaluated using CSI,

\begin{equation}
\mathrm{CSI}
=
\frac{H}{H+M+FA},
\end{equation}

where $H$, $M$, and $FA$ denote hits, misses, and false alarms, respectively. 

Neighborhood spatial skill was evaluated using FSS. Forecast and observed precipitation fields were converted to binary threshold-exceedance fields, and exceedance fractions were calculated within a $15\times15$ grid-cell neighborhood. This neighborhood allows modest spatial displacement between forecast and observed precipitation while reducing the double-penalty associated with pointwise verification. FSS was calculated as

\begin{equation}
\mathrm{FSS}
=
1-
\frac{
\sum_{i=1}^{N}
\left(f_i-o_i\right)^2
}{
\sum_{i=1}^{N} f_i^2
+
\sum_{i=1}^{N} o_i^2
},
\end{equation}

where $f_i$ and $o_i$ denote the forecast and observed neighborhood exceedance fractions, respectively. FSS ranges from 0 to 1, with values approaching 1 indicating greater spatial agreement. 

Differences between forecast and observed distributions of 24-h precipitation accumulations were quantified using a histogram-based PDF distance. Precipitation values were divided into 40 equally spaced bins spanning 0--20~mm, corresponding to a bin width of 0.5~mm, and converted to relative frequencies,

\begin{equation}
p_j = \frac{n_j}{\sum_{k=1}^{B} n_k},
\end{equation}

where $n_j$ is the number of grid points within bin $j$ and $B=40$. The PDF distance was calculated as

\begin{equation}
D_{\mathrm{PDF}}
=
\sqrt{
\frac{1}{B}
\sum_{j=1}^{B}
\left(
p_j^{F}-p_j^{O}
\right)^2
},
\end{equation}

where $p_j^{F}$ and $p_j^{O}$ denote the forecast and observed relative frequencies, respectively. Smaller values indicate greater agreement between the forecast and observed precipitation distributions. The 0--20~mm range was used to characterize the main precipitation distribution while maintaining sufficient sampling within individual bins, whereas heavier precipitation was evaluated separately using the percentile-based metrics.

\subsection{Statistical analysis}

Verification metrics were first calculated independently for each 24-h AR evaluation window and subsequently averaged across the corresponding set of windows. Each evaluation window therefore contributed equally to the aggregate statistics irrespective of the number of grid points within its evaluation domain, preventing larger AR domains from contributing disproportionately to the reported model performance.

Uncertainty in the aggregate metrics was quantified using an event-level clustered bootstrap to account for dependence among overlapping evaluation windows originating from the same continuous AR event. For each bootstrap replicate, the same number of continuous AR events as in the original sample was sampled with replacement, and all 24-h evaluation windows associated with each selected event were retained together. The aggregate metric was then calculated across the resulting windows using the same equal-window averaging procedure as in the primary analysis. This procedure was repeated 5,000 times, and the 2.5th and 97.5th percentiles of the resulting distribution were used to define the 95\% confidence interval. A fixed random seed of 42 was used for reproducibility.

\section{Conclusion}\label{sec_conclusion}
This study provides a global evaluation of AR-related precipitation forecasts from GFS, GEFS Mean, GraphCast, and AIFS from Day 1 through Day 10. Across a globally distributed collection of AR events, GraphCast and AIFS consistently exhibit greater spatial skill than the NWP systems, with the largest differences occurring for heavy precipitation and at longer forecast lead times. These improvements persist across North America, Europe, and Australia and New Zealand and across seasons, indicating that the spatial advantage of AIWP is not limited to a particular geographical region or time of year.

Greater spatial skill, however, does not translate directly into more accurate precipitation magnitude. GraphCast and AIFS tend to overpredict moderate-to-heavy precipitation while underpredicting the heaviest observed accumulations at longer lead times, whereas GFS and GEFS Mean develop pronounced dry biases. Representative AR events further illustrate this distinction, with the AIWP models better preserving the broad location and organization of AR-related precipitation while not consistently reproducing localized precipitation maxima. These results demonstrate that spatial skill and precipitation magnitude represent distinct aspects of forecast quality that should be evaluated separately when assessing AIWP systems for high-impact precipitation.

Overall, AIWP models provide substantial improvements in the spatial prediction of AR-related precipitation, particularly for heavy precipitation and at longer lead times, while accurate representation of precipitation magnitude and localized extremes remains an important challenge. The standardized event-based evaluation implemented within the Extreme Weather Bench framework provides a reproducible basis for assessing whether future advances in AI and physics-based forecasting systems translate into improvements across both dimensions of high-impact precipitation prediction.

\backmatter


\bmhead{Acknowledgements}
This material is based upon work supported in part by Brightband and by the National Science Foundation under Grant No. RISE-2019758 (NSF AI Institute for Research on Trustworthy AI in Weather, Climate, and Coastal Oceanography, AI2ES), and under Grant No. IIS2324008. Any opinions, findings, and conclusions or recommendations expressed in this material are those of the author(s) and do not necessarily reflect the views of Brightband or the National Science Foundation.

\bmhead{Author Contributions}
M.V.: Conceptualization, Methodology, Software, Formal analysis, Investigation, Data curation, Visualization, Validation, Writing -- original draft, Writing -- review and editing.

T.M.: Conceptualization, Methodology, Software, Investigation, Data curation, Validation, Writing -- review and editing.

A.M.: Conceptualization, Methodology, Investigation, Validation, Supervision, Project administration, Resources, Funding acquisition, Writing -- review and editing.

A.H.: Conceptualization, Methodology, Investigation, Validation, Supervision, Project administration, Writing -- review and editing.

D.R.: Data curation, Validation, Resources, Funding acquisition, Writing -- review and editing.

All authors reviewed and approved the final manuscript.

\bmhead{Data Availability}
GFS and GEFS data used in this study are publicly available through the National Oceanic and Atmospheric Administration (NOAA) Open Data Dissemination program via the Registry of Open Data on AWS. GFS data are available at \url{https://registry.opendata.aws/noaa-gfs-bdp-pds/}, and GEFS data are available at \url{https://registry.opendata.aws/noaa-gefs/}. GraphCast and AIFS datasets were obtained from Brightband and available for use under the CC-BY-SA 4.0 license; they can be retrieved from the Earthmover Marketplace at \url{https://app.earthmover.io/marketplace/6969541cb7bd57a837d648c8} or from Google Cloud Storage at \url{gs://brightband-public-mlwp-forecast-archive}. The Extreme Weather Bench framework used in this study was introduced by \citep{mcgovern2026extreme} and is publicly available at \url{https://github.com/brightbandtech/ExtremeWeatherBench}; version 1.0.2 was used in this study and is archived on Zenodo at \url{https://zenodo.org/records/19924742}.

\bmhead{Code Availability}
The code used to process the datasets, perform the precipitation forecast evaluation, compute the verification metrics, and generate the figures presented in this study is publicly available on GitHub at \url{https://github.com/ai2es/ai_nwp_ar_precipitation}.

\noindent






\begin{appendices}
\appendix

\section{Bootstrap Confidence Intervals}
\label{appendix_bootstrap_ci}
Event-level clustered bootstrap analysis was used to quantify uncertainty in the aggregate verification metrics, as described in Section~\ref{sec_methods}. Appendix Figs.~1--4 show the resulting 95\% confidence intervals for bias, Pearson correlation, FSS, and CSI at Days~1, 5, and~10 across North America, Europe, and Australia and New Zealand. The confidence intervals are generally narrow relative to both the differences among forecasting systems and the ranges of the corresponding metrics, although greater uncertainty occurs for selected metrics and models at longer lead times. Importantly, the confidence intervals remain narrow after accounting for the dependence among overlapping 24-h evaluation windows through event-level resampling, indicating that the aggregate differences among forecasting systems are robust to sampling variability among AR events. The relatively narrow intervals are consistent with the large number of independent continuous AR events included in the evaluation. Each evaluation-window metric also summarizes precipitation performance across the full AR evaluation domain, providing a spatially aggregated measure of forecast performance rather than relying on individual grid-point errors.

\begin{figure}[!htbp]
    \centering
    \includegraphics[width=\textwidth]
    {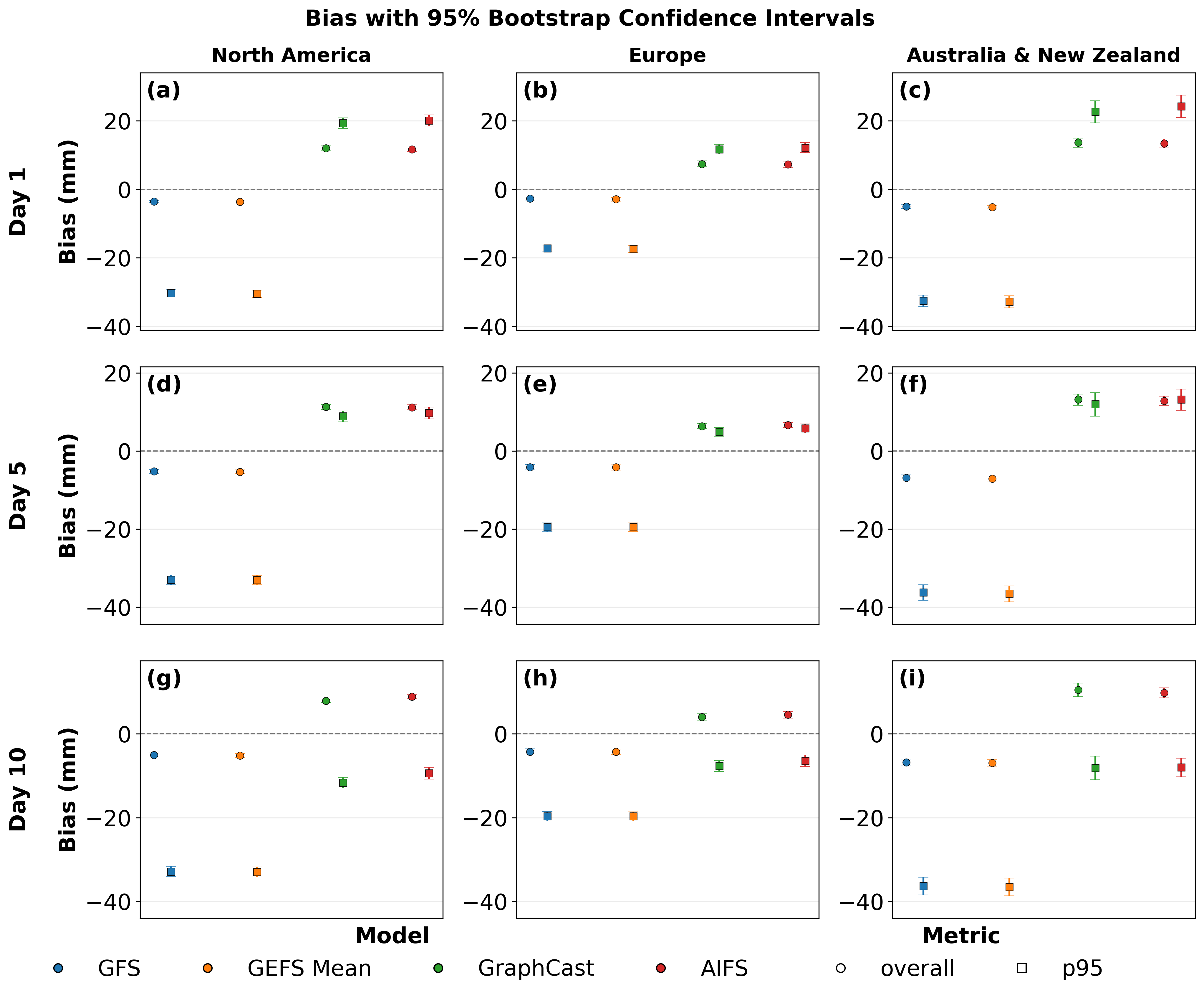}
    \caption{\textbf{Regional precipitation bias with 95\% bootstrap confidence intervals.}
    Mean bias for GFS, GEFS Mean, GraphCast, and AIFS at Day 1
    (\textbf{a--c}), Day 5 (\textbf{d--f}), and Day 10 (\textbf{g--i}).
    Columns correspond to North America (\textbf{a,d,g}), Europe
    (\textbf{b,e,h}), and Australia and New Zealand (\textbf{c,f,i}).
    Circles show overall bias and squares show bias conditioned on observed
    precipitation exceeding the local 95th-percentile threshold. Error bars
    indicate 95\% confidence intervals obtained from 5,000 event-level
    clustered bootstrap replicates.}
    \label{fig:supp_bias_ci}
\end{figure}

\begin{figure}[!htbp]
    \centering
    \includegraphics[width=\textwidth]
    {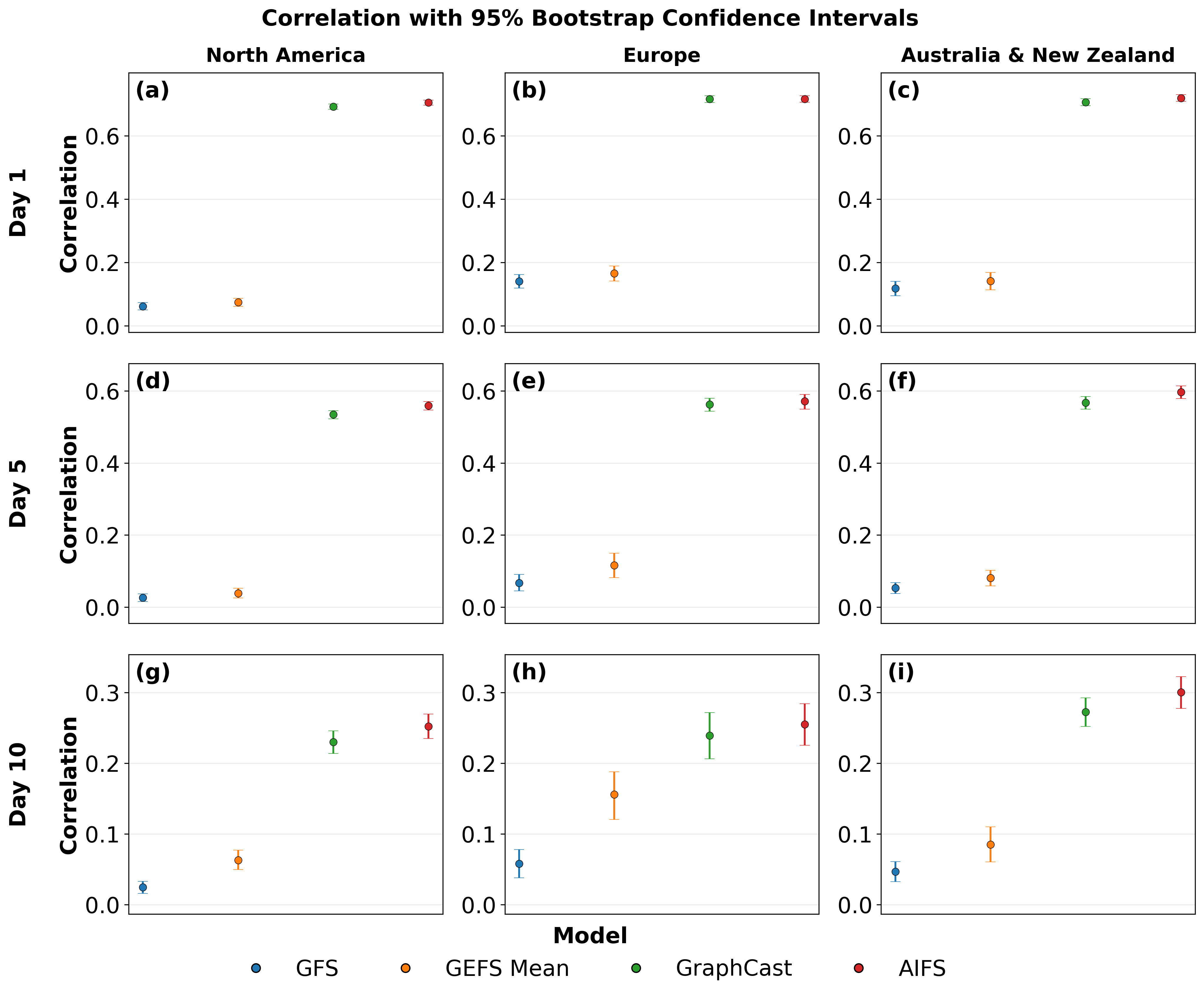}
    \caption{\textbf{Regional Pearson correlation with 95\% bootstrap confidence intervals.}
    Mean Pearson correlation for GFS, GEFS Mean, GraphCast, and AIFS at
    Day 1 (\textbf{a--c}), Day 5 (\textbf{d--f}), and Day 10
    (\textbf{g--i}). Columns correspond to North America
    (\textbf{a,d,g}), Europe (\textbf{b,e,h}), and Australia and New Zealand
    (\textbf{c,f,i}). Error bars indicate 95\% confidence intervals obtained
    from 5,000 event-level clustered bootstrap replicates.}
    \label{fig:supp_correlation_ci}
\end{figure}

\begin{figure}[!htbp]
    \centering
    \includegraphics[width=\textwidth]
    {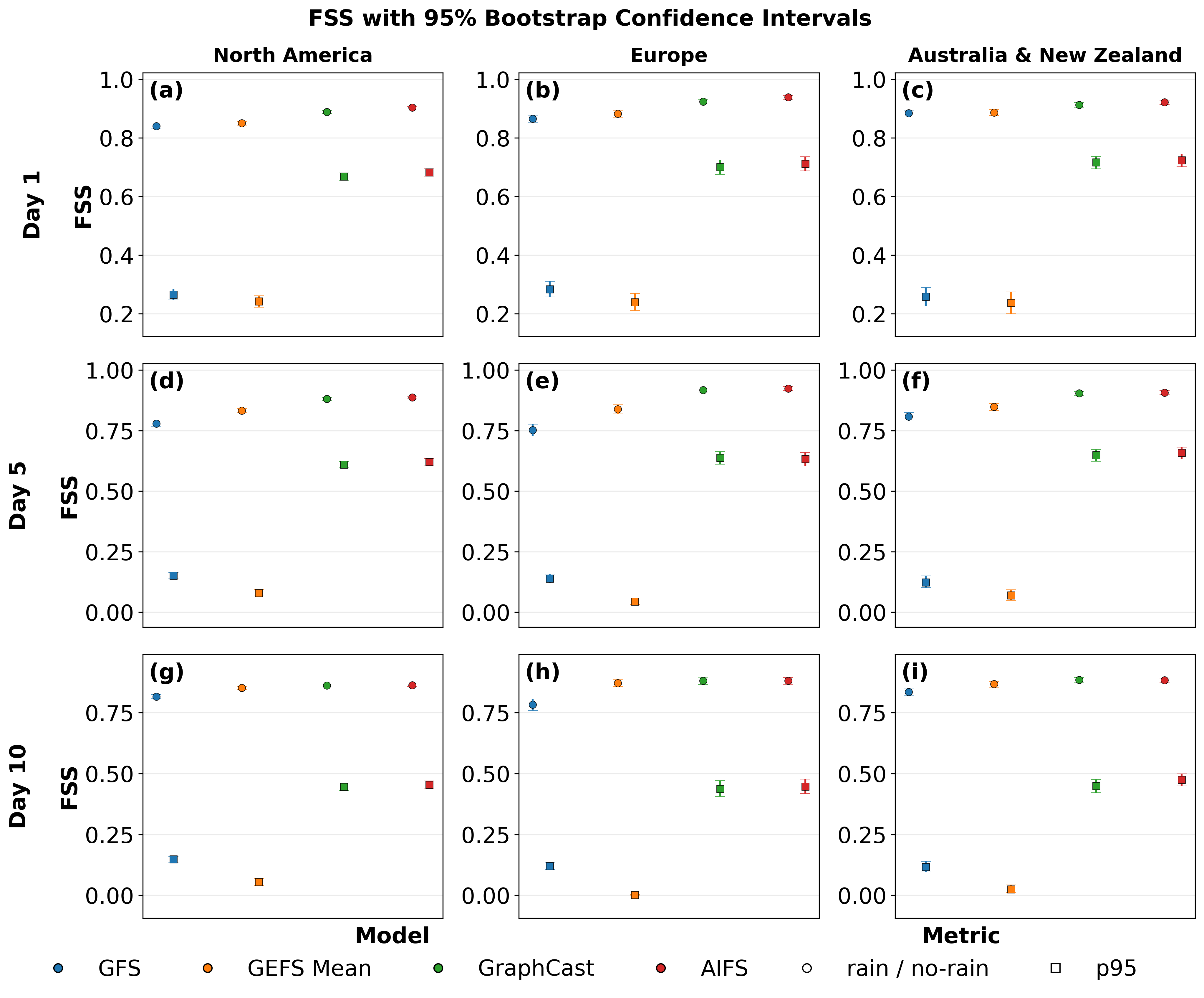}
    \caption{\textbf{Regional Fractions Skill Score with 95\% bootstrap confidence intervals.}
    Mean Fractions Skill Score (FSS) for GFS, GEFS Mean, GraphCast, and AIFS
    at Day 1 (\textbf{a--c}), Day 5 (\textbf{d--f}), and Day 10
    (\textbf{g--i}). Columns correspond to North America
    (\textbf{a,d,g}), Europe (\textbf{b,e,h}), and Australia and New Zealand
    (\textbf{c,f,i}). Circles show rain/no-rain FSS and squares show FSS
    evaluated at the local 95th-percentile precipitation threshold. Error
    bars indicate 95\% confidence intervals obtained from 5,000 event-level
    clustered bootstrap replicates.}
    \label{fig:supp_fss_ci}
\end{figure}

\begin{figure}[!htbp]
    \centering
    \includegraphics[width=\textwidth]
    {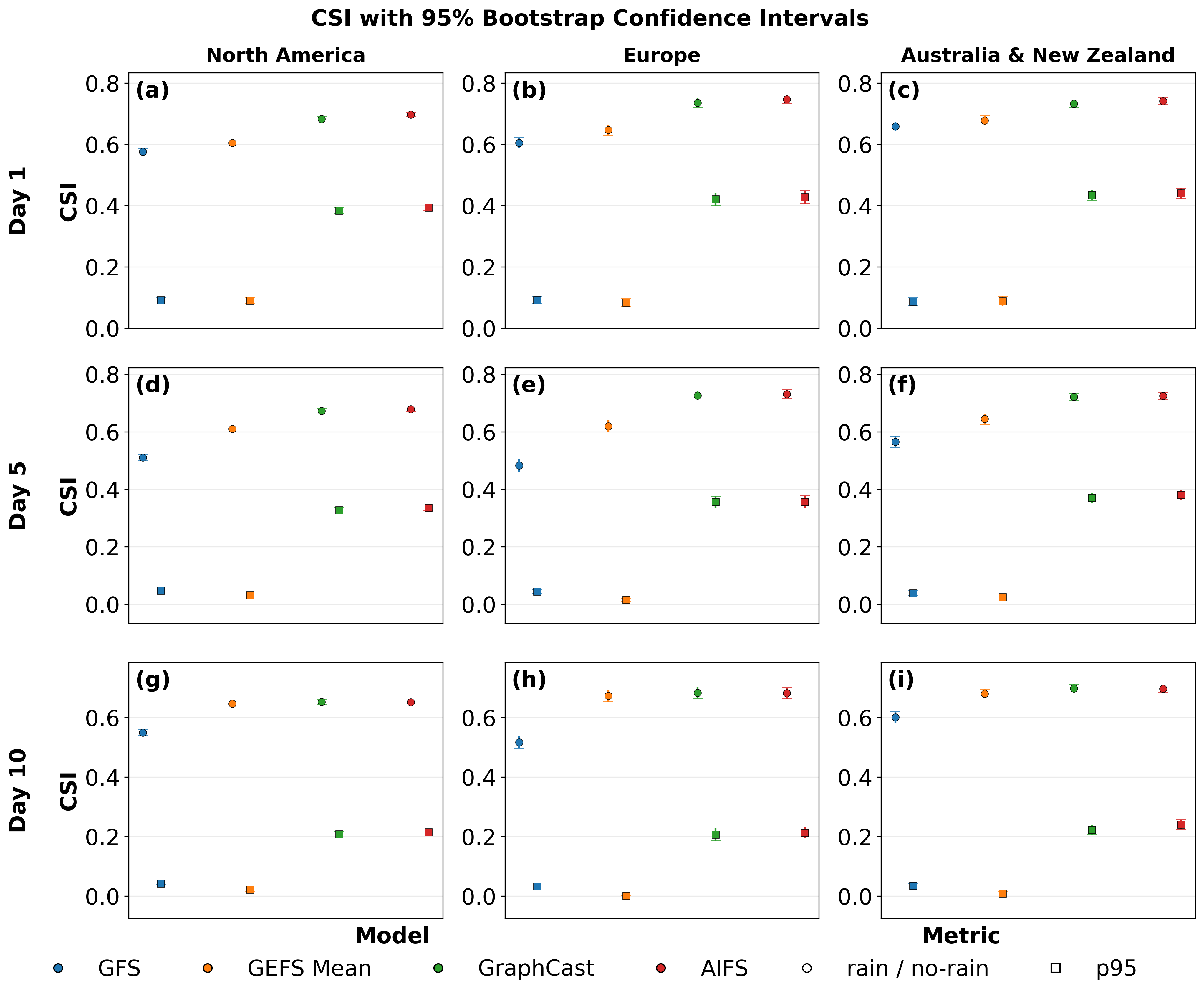}
    \caption{\textbf{Regional Critical Success Index with 95\% bootstrap confidence intervals.}
    Mean Critical Success Index (CSI) for GFS, GEFS Mean, GraphCast, and AIFS
    at Day 1 (\textbf{a--c}), Day 5 (\textbf{d--f}), and Day 10
    (\textbf{g--i}). Columns correspond to North America
    (\textbf{a,d,g}), Europe (\textbf{b,e,h}), and Australia and New Zealand
    (\textbf{c,f,i}). Circles show rain/no-rain CSI and squares show CSI
    evaluated at the local 95th-percentile precipitation threshold. Error
    bars indicate 95\% confidence intervals obtained from 5,000 event-level
    clustered bootstrap replicates.}
    \label{fig:supp_csi_ci}
\end{figure}

\clearpage




\end{appendices}


\bibliography{references}

\end{document}